\documentclass[nohyper,notoc]{JHEP}

\usepackage{amsmath,euscript,array,amssymb,cite,bm} 
\def\St{{\tilde{S}}}
\def\b{{\beta}}
\def\a{{\alpha}}
\def\m{{\mu}}
\def\th{{\theta}}
\def\gh{{\hat\gamma}}
\def\l{{\lambda}}

\def\cN{{\cal N}}

\def\Dbarslash{\,\,{\raise.15ex\hbox{/}\mkern-12mu {\bar\D}}}
\def\Dslash{\,\,{\raise.15ex\hbox{/}\mkern-12mu \D}}
\def\delslash{\,\,{\raise.15ex\hbox{/}\mkern-9mu \partial}}
\def\delbarslash{\,\,{\raise.15ex\hbox{/}\mkern-9mu {\bar\partial}}}

\def\hf{{\textstyle{1\over2}}}

\def\l{\lambda}

\def\D{{\cal D}}
\def\Dbarslash{\,\,{\raise.15ex\hbox{/}\mkern-12mu {\bar\D}}}
\def\delslash{\,\,{\raise.15ex\hbox{/}\mkern-9mu \partial}}
\def\Dslash{\,\,{\raise.15ex\hbox{/}\mkern-12mu \D}}

\def\adss{AdS_5\times S^5}

\newcommand{\AdS}{AdS}

\def\bigR{{\rm I}\!{\rm R}}

\def\={\, =\, }
\def\+{\, +\, }
\def\-{\, -\, }

\newcommand{\be}{\begin{equation}}
\newcommand{\ee}{\end{equation}}
\def\bea{\begin{eqnarray}}
\def\eea{\end{eqnarray}}

\title{ Magnons, Classical Strings and $\bm{\beta }$-Deformations}

\author{Chong-Sun Chu$^a$, George Georgiou$^b$ and Valentin V.~Khoze$^b$\\
$^a$Department of Mathematical Sciences, University of Durham,
Durham, DH1 3LE, UK\\
$^b$Department of Physics and IPPP, University of Durham,
Durham, DH1 3LE, UK\\
{\tt chong-sun.chu@durham.ac.uk,}\, 
{\tt george.georgiou@durham.ac.uk,}\, 
{\tt valya.khoze@durham.ac.uk}}

\abstract{Motivated by the recent work of Hofman and Maldacena [1] we construct
a classical string solution on the $\beta$-deformed $AdS_5 \times \tilde{S}^5$ background.
This string solution is identified with a magnon state
of the integrable spin chain description
of the $\cN=1$ supersymmetric $\beta$-deformed gauge theory. 
The string solution carries two angular momenta, an infinite $J_1$ and 
a finite $J_2$ which classically can take arbitrary values.
This string solution corresponds to the magnon of charge $J_2$ propagating on an
infinite spin chain. 
We derive an exact dispersion relation for this magnon from string theory.}
 
\preprint{{\tt hep-th/0606220}
\\IPPP/06/37\\
DCPT/06/74
}

\begin{document}

\section{Introduction}

The AdS/CFT correspondence states the equivalence of string theory on
$\adss$ to the $\cN=4$ supersymmetric Yang-Mills \cite{adscft}. One
consequence of this duality is that the spectrum of string states must
match with the spectrum of operator dimensions in gauge theory.
This statement has been tested initially only for supergravity
multiplets and their KK descendants. Since quantization of string theory on $\adss$
is not fully understood, a more complete verification of the spectral matching
has posed substantial difficulties.

A significant new step was made in
\cite{BMN,GKP} based on the identification of a particular sector
of string states carrying a large angular momentum $J$ with 
the `long' gauge theory operators.
The authors of
\cite{GKP} have argued that the spectrum of the large-$J$ string states
can be computed reliably in the semi-classical
approximation. 

On the gauge theory side the problem of determining the spectrum corresponds to diagonalizing
the dilatation operator. In planar perturbation theory, for small 't Hooft coupling $\lambda=Ng^2 \ll 1$
this problem has an elegant reformulation \cite{chain1,chain2,Dippel,AFSTA,chainrev} 
in terms of diagonalizing an integrable spin chain. 
On the string theory side, the problem simplifies in the opposite limit
of $\lambda \gg 1,$ where the string sigma-model becomes weakly coupled. It has been argued
in \cite{Str1,Str2} that the classical string sigma-model on the $\adss$ is also integrable.
The appearance of integrability on both sides of the correspondence (albeit in opposite limits)
has triggered a lot of interest and offered a hope that the prediction of the matching of the spectra
can be tested and verified explicitly. 

In the integrable spin chain description of gauge theory, the Yang-Mills
composite operators are assembled from specific  building blocks
which are associated with magnons -- the elementary excitations of the spin chain with one
flipped spin.
The Yang-Mills description of the magnon corresponds to the
operator 
\be
{\cal O} \, \sim \, \sum_l \, e^{ipl}\, (\cdots ZZZ\, I \, ZZZ \cdots)
\label{HMmagn}
\ee 
Here the `impurity' $I$ is inserted at a position $l$ along the chain of 
$Z$ fields ($J$ of them). 
In the simplest settings which correspond to the $SU(2)$ sector
of the $\cN=4$ gauge theory, the $Z$ field is given by a complex scalar field
$\Phi_1,$ and the impurity $I$ is given by another complex scalar $\Phi_2.$ 

Each magnon is characterized by its dispersion relation,
\be \label{disp}
E-J = \sqrt{1+ \frac{\l}{\pi^2} \sin^2 \frac{p}{2}}
\ee
In the limit of a long spin chain, $J \gg \infty,$ the magnons are dilute. 
As a result, the scaling
dimensions of Yang-Mills operators can be computed by summing over the
dispersion relations \eqref{disp} for each constituent magnon.
Equation \eqref{disp}
is an exact BPS formula\footnote{More precisely, 
the dispersion relation is of the form $E-J = \sqrt{1+ f(\l) \sin^2{p}/{2}}.$
Supersymmetry alone cannot determine the function $f(\l).$ However, all known
perturbative (up to 3-loops) and the strong-coupling results are consistent with
$f(\l)=\l/\pi^2.$ } 
\cite{beisert-susy} in the $\cN=4$ gauge theory. It can be
derived from the supersymmetry algebra
\cite{beisert-susy}, or by adopting the calculation in \cite{sz}. 

The AdS/CFT correspondence relates gauge invariant Yang-Mills operators to 
string states. In the integrable
spin chain description, these operators are assembled from 
magnons. It is natural to ask if the magnon itself
has a string description. In a recent paper \cite{HM} Hofman and Maldacena 
have found this description.

Hofman and Maldacena \cite{HM} have considered a particular double-scaling $N \to \infty$
limit where 
\be
J \to \infty \ , \qquad \lambda = {\rm fixed} \ , \qquad
p = {\rm fixed} \ , \qquad E-J = {\rm fixed}
\ee
In this limit
both, the spin chain of the $\cN=4$ gauge theory, and the classical string 
on the $\adss$, become infinitely long. In this sector the problem of determining the spectra
of both theories becomes more tractable. The spectrum of the infinite spin chain can be constructed
in terms of asymptotic states. These asymptotic states are made out of elementary excitations of the spin chain 
-- the magnons. They carry a conserved momentum $p$ and have the energy $\epsilon (p)=E-J-1$ given by
Eq.~\eqref{disp}.
On the string theory side,
the authors of Ref.~\cite{HM} have found a classical solution which precisely corresponds to an
elementary magnon of the gauge theory spin chain. The lightcone energy $E-J$
of this classical string coincides with the dispersion relation of the magnon \eqref{disp}.

In the subsequent work \cite{Dorey} N. Dorey et al have constructed classical string solutions
which correspond to bound states of magnons of the spin chain.
There is an exact dispersion relation \cite{Dorey}
which holds for the spin chain magnons of charge $J_2$
and for the semi-classical string with angular momenta $J_1$ and $J_2$:
\be
E-J_1 \,  =\, \sqrt{J_2^2 \, +\, \frac{\lambda}{\pi^2} \sin^2 \frac{p}{2}}
\label{disp2}
\ee
These results of \cite{HM,Dorey}
imply that in the infinite spin chain limit, the $\cN=4$ SYM spectrum represented
by elementary and composite magnon asymptotic sates matches precisely with the states in 
semi-classical string 
theory on $\adss.$
Two very recent papers \cite{AFZ,MTT} analyze  the effects of finite $J$ and of the
quantum corrections to the semi-classical string magnons.

The main motivation of this paper is to study what happens with this
classical-string/magnon correspondence in theories with less
supersymmetry. We will consider an $\cN=1$ supersymmetric gauge theory
obtained by a marginal $\beta$-deformation of the $\cN=4$ SYM. The
AdS/CFT duality extends to the $\beta$-deformed theories where it
relates the $\cN=1$ $\beta$-SYM and the supergravity on the deformed
$\AdS_5\times \tilde{S}^5$ background. The gravity dual was found by
Lunin and Maldacena in Ref.~\cite{LM}.
On the other hand, for real $\beta$, 
the $\beta$-SYM theory has an integrable spin chain description 
\cite{Roiban,Cherkis,FRT,BR} and also the
Lax pair exists \cite{Frolov}.
Hence, it is interesting to find out if the classical-string/magnon correspondence
holds in the  $\beta$-deformed case.

In \cite{FRT}, it was argued that the two-loop $SU(2)$ spin chain Hamiltonian representing the 
two-loop dilation operator of the real-$\beta$-deformed SYM can be obtained from the 2-loop 
spin chain Hamiltonian of the undeformed ${\cal N}=4$ theory by applying a position-dependent unitary operator ${\cal U}$. 
The effect of this unitary operator is to twist the boundary conditions of the original spin chain. 
As a result, the asymptotic Bethe ansatz equations for the deformed theory can be obtained from the  
Bethe ansatz equations of the undeformed theory simply by performing the following substitution 
\be
\label{p-b}
p \longrightarrow p-2\pi \beta
\ee  
on the left-hand side of the Bethe ansatz equations. 
If this construction of the spin chain Hamiltonian of the deformed theory can be extended 
to higher loops then one can calculate the anomalous dimensions of all single trace operators 
in the $SU(2)$ sector of the deformed theory by taking the conjectured all-loop Bethe equations 
for the undeformed theory \cite{Dippel,AFSTA} and modifying it as stated in \eqref{p-b}.
Hence for real-valued deformation parameter $\beta \in \bigR$, 
we expect that
the dispersion relation \eqref{disp2} is modified only through a shift
in the momentum $p$:
\be
E-J_1 \,  =\, \sqrt{J_2^2 \, +\, \frac{\lambda}{\pi^2} \sin^2 
\left(\frac{p}{2}-\pi \b\right)}
\label{disp3}
\ee
This modification can 
also be traced back to a perturbative calculation of
anomalous dimensions of operators assembled from \eqref{HMmagn} in the
$\beta$-deformed theory.

Since in the Hofman-Maldacena limit the magnon momentum
$p$ is kept fixed and that for the LM supergravity solution to hold,
$\beta$ must be small, $\b \ll 1/R \ll 1,$ it follows that the
dispersion relations for the magnon and multi-magnon states of the
$\beta$-deformed theory are the same as those in the undeformed case
\eqref{disp2}. 

If the Hofman-Maldacena construction does extend to the $\beta$-deformed
AdS/CFT correspondence, the magnon of the spin chain must correspond to
(an open part of) a fundamental string moving on the deformed sphere
$\St^5.$ Since the background itself depends on the deformation
parameter $\gh,$ one would expect that the relevant classical string
solution will carry a lightcone energy $E-J$ that depends nontrivially
on the deformation. We will find that this is indeed the case for the
$\beta$-deformed solution constructed in this paper. From this one might
expect that the dispersion relation for magnons of the $\beta$-deformed
theory would explicitly depend on the deformation parameter $\gh= \b
\sqrt{\l}.$ If this was true, this would be in contradiction with the
integrable description of the $\beta$-deformed theory.

It will turn out that the entire $\gh$-dependence in the dispersion 
relation for the string is absorbed into the second angular momentum $J_2.$ 
In the regime where the Lunin-Maldacena supergravity background is reliable
($\beta \ll 1,$ $\gh$-fixed) our solution will precisely reproduce the dispersion relation
\eqref{disp2}. 
The $\beta$-deformed classical open string solution corresponds to a magnon of charge $J_2$
in the spin chain description of the $\beta$-deformed gauge theory. This magnon is 
schematically\footnote{More precisely, as explained in \cite{Dorey}, the magnon bound-state
corresponds to a particular state of the spin chain where the wave-function is strongly
peaked on configurations where all flipped spins (i.e. $\Phi_2$'s) are nearly adjacent to 
each other.}
of the form
\eqref{HMmagn} where $Z=\Phi_1$ and the impurity is composite $I \sim (\Phi_2)^{J_2}.$
 
In this paper we will consider only deformations with $\b$ real.
In the undeformed $\cN=4$ theory
the formula \eqref{disp2} is the dispersion relation for states 
in a short representation. Because it is BPS protected in the $\cN=4$ theory, it is
valid for any values of $\lambda.$ We will re-derive the same formula in the $\cN=1$ theory.
In this case we cannot appeal to the BPS properties based on the extended supersymmetry algebra.  
One particularly pleasing feature of our analysis is that we will be able to derive the 
full square-root expression in \eqref{disp2} from classical string theory.
Based on expectations from the integrable spin chain analysis 
it is likely that the dispersion relation \eqref{disp2} is exact and is valid for
all values of $\l$ in the $\beta$-deformed $\cN=1$ theory.

Recent papers which study perturbative and non-perturbative effects
in $\beta$-deformed gauge theories include
\cite{FG,PSZ,Rossi,MPSZ,VVK,GK,CK,MPPSZ,DGK,EMPSZ}.

%
\section{Classical strings in the $\beta$-deformed background}
%


The supergravity background 
dual to $\beta$-deformed gauge theory
was constructed by Lunin and Maldacena (LM) \cite{LM} by applying a solution generating
$SL(3,R)$ transformation to the $\adss$ background, or equivalently 
an STsTS${}^{-1}$ transformation.   
The deformed supergravity solution \cite{LM} contains
the metric on $AdS_5 \times \St^5$
\be
\label{metricdef}
ds^2_{str} \,=\, R^2\, \left[ ds^2_{AdS_5} +  
\sum_{i=1}^3 ( d\mu_i^2  + G \mu_i^2 d\phi_i^2) 
+ \gh^2 \,G \mu_1^2\mu_2^2\mu_3^2 
(\sum_{i=1}^3 d\phi_i)^2 \right] 
\ee
where $\St^5$ is a $\beta$-deformed five-sphere and 
$\sum_{i=1}^3 \mu_i^2=1.$
The LM solution
also involves the dilaton-axion field $\tau$ as well as the 
RR and NS-NS form fields. In what follows we will require only the expression for the
metric \eqref{metricdef} and the
the NS-NS two-form field
\be
\label{B2def}
B^{\rm NS}_2 =\,
\gh R^2 G\, (\m_1^2 \m_2^2 d\phi_1 d\phi_2 
+ \m_2^2 \m_3^2 d\phi_2 d\phi_3
+ \m_3^2 \m_1^2 d\phi_3 d\phi_1)
\ee
Here 
\be \label{Gdef}
G^{-1}=\, 1+ \gh^2 (\m_1^2 \m_2^2 + \m_2^2 \m_3^2 + \m_1^2 \m_3^2)\ , 
\qquad R^4 :=\, 4 \pi N g_{\rm st}\ ,
\qquad \gh :=\, R^2 \b \,=\, \sqrt{\lambda}\b \ .
\ee
 
The coordinates $(\mu_i,\phi_i)$ which parameterize the deformed 5-sphere $\St^5$
correspond precisely to the three complex scalars $\Phi_i$ of the $\beta$-deformed gauge theory.
This correspondence is dictated by the three $U(1)$ isometries surviving from the $SU(4)_R$ symmetry
of the $\cN=4$ SYM,
\bea
\mu_1 \, e^{i\phi_1} &=& \Phi_1 \,=\, \varphi^1\,+\, i \varphi^2 \ , \label{mu1phi1} \\
\mu_2 \, e^{i\phi_2} &=& \Phi_2 \,=\, \varphi^3\,+\, i \varphi^4 \ , \label{mu2phi2} \\
\mu_3 \, e^{i\phi_3} &=& \Phi_3 \,=\, \varphi^5\,+\, i \varphi^6 \ , \label{mu3phi3} 
\eea
Here $\Phi_i$ denote complex scalars which are the lowest components of the three chiral
superfields of the $\cN=1$ supersymmetric $\beta$-deformed gauge theory and
$\varphi^1, \ldots, \varphi^6$ denote the corresponding six real scalar fields.
It will be convenient to parameterize  $\m_i$ coordinates via
\be
\m_3= \sin \th \sin \a\ , \qquad
\m_1 = \sin \th \cos \a\ , \qquad
\m_2 = \cos \th 
\label{mugen}
\ee
so that $\sum_{i} d \m_i^2 = d \th^2 + \sin^2 \th \; d \a^2$. 

As mentioned earlier,
the LM supergravity background is a reliable approximation to string theory
in the regime \cite{LM} where
$R \gg 1$ and  $\b \ll 1 .$

Ultimately we are interested in closed bosonic strings moving on $\bigR \times \St^5$.
These closed (folded) strings can be constructed from open strings in the same manner
as in \cite{HM}. It is an open string with the ends on the equator of the deformed sphere 
which corresponds to a magnon building block of the gauge theory operators. From now on we concentrate on 
such open strings.

\underline{$\gh =0$: Hofman-Maldacena solution}

First we briefly recall the classical string solution in the undeformed $\cN=4$
theory found by Hofman and Maldacena in \cite{HM}. The deformations are switched off
by setting $\gh=0$ (which also implies $G=1$ with $B^{\rm NS}=0$) in 
Eqs.~\eqref{metricdef}-\eqref{Gdef} above. 
This solution 
lives on the $\bigR \times S^2$ background, where the metric on $S^2$ is
\be
ds^2\, =\, d \theta^2 \,+ \, \sin^2 \theta \, d\phi^2
\label{s2metr}
\ee

We are looking for a 
solution of equations of motion arising from the Polyakov action in the conformal gauge.
It corresponds to a classical open string moving on the infinite worldsheet
parameterized by coordinates $t$ and $x.$

This solution $\theta(x,t),$ $\phi(x,t)$ we are after can be written in the form
\be
\theta(x,t) \, =\, \theta(y) \ , \qquad
\phi(x,t) \, =\, t \, +\,  g(y) \ , \qquad 
y\, :=\, cx\, -\, dt
\label{HMans}
\ee
where $c$ and $d$ are positive constants. Functions $\theta(y)$ and $g(y)$ represent a wave
localized around $y=0$ and
moving with a group velocity $v=d/c \le 1.$ Apart from the $y$-dependence, the angle $\phi$ in \eqref{HMans}
also depends on $t$ linearly. This is interpreted as a string rotating in the azimuthal $\phi$-direction 
and gives rise to a large
angular momentum $J=\partial S/\partial {\dot\phi}.$
The explicit form of the HM solution reads \cite{HM}
\be
\cos \theta \,=\, \frac{1}{c} \, \frac{1}{{\rm cosh}\, y} \ , \qquad
\tan g \,=\, \frac{1}{d} \, {\rm tanh}\, y
\label{HMansw}
\ee
where the constants $c$ and $d$ are given by
\be
c \, =\, \frac{1}{\cos \theta_0} \ , \qquad d\, =\, \tan \theta_0 \ , \qquad c^2 -d^2 =1
\ee
and $\cos \theta_0$ is the maximal value of $\cos \theta$ in \eqref{HMansw}.
This solution is characterized by two integrals of motion,
the energy $E,$ and the $\phi$-angular momentum $J.$ Both $E$ and $J$ are infinite
quantities when evaluated on the solution, but
the combination $E-J$ is finite,
\be
E-J \, =\, \frac{\sqrt{\lambda}}{\pi}\, \cos \theta_0
\label{disprel1}
\ee

This classical string solution corresponds to an elementary magnon in the integrable 
spin chain description of the $\cN=4$ gauge theory. 
Let us choose the azimuthal angle on the $S^2$ sphere \eqref{s2metr} to be $\phi_1$ for concreteness.
This is achieved by setting $\alpha=0$ on the $S^5$ so that
\be
\m_1 = \sin \th \ , \qquad
\m_2 = \cos \th \ , \qquad
\m_3= 0
\label{mus3}
\ee
In addition we decouple $\phi_2$ and $\phi_3$ by setting them to zero.
The resulting metric on $S^2$ is given by \eqref{s2metr} with $\phi=\phi_1.$
In the co-moving frame (i.e. in terms of the $(t,y)$ coordinates)
the string solution is represented by a time-independent $\theta(y)$ and a 
time-dependent $\phi_1=t+g_1(y).$ This corresponds to a string rotating
in the azimuthal $\phi_1$ direction. In terms of the $\mu_i$ coordinates
\eqref{mus3} the rotation is around $\mu_1$
\be
\m_1 \,e^{i\phi_1(t)}= \sin \th\,e^{i\phi_1(t)}  \ , \qquad 
\m_2\, = \,\cos \th \ , \qquad \m_3 \,=\, 0 \ , \qquad
\th = {\rm const}
\ee
Comparing this with the gauge theory dictionary \eqref{mu1phi1}-\eqref{mu3phi3},
we see that in this case the Hofman-Maldacena string solution corresponds to a magnon
\eqref{HMmagn} with $Z=\Phi_1$ and the impurity $I=\Phi_2.$
The exact
dispersion relation for this magnon is given by
\be
E-J_1 \, =\, \sqrt{1+ \frac{\lambda}{\pi^2} \sin^2 \frac{p}{2}}
\ee
which in the large $\lambda$ limit coincides with the classical string result
\eqref{disprel1} provided that one makes an identification \cite{HM}
\be
\sin \frac{p}{2}\, =\, \cos \theta_0
\label{HMident}
\ee


\underline{$\gh \neq 0$: String solution parameterized by two angles}

The main motivation behind this paper is to find what happens to the Hofman-Maldacena solution
when one switches on the deformation parameter $\gh.$
The most obvious thing seems to be to construct the appropriate solution 
in terms of $\theta(y)$ and $\phi_1(t,y)$ on the $\beta$-deformed background.
This will be done in the next sub-section where we will see  that the minimal
such solution will necessarily involve the third angle, e.g. $\phi_2$
and will be forced to propagate on $\St^3$ rather than $\St^2.$

Before turning to this case we want to comment on a more trivial case
of the solution propagating on a 2-sphere which does not involve $\phi_i$'s, but instead
is parameterized by the 
angles $\theta$ and $\alpha$.
For simplicity we set all $\phi_i$ angles to zero. At this point the
deformed $\St^5$ sphere collapses to the ordinary $S^2$ sphere 
with the metric
\be
ds^2\, =\, d \theta^2 \,+ \, \sin^2 \theta \, d\alpha^2
\label{s2metr2}
\ee
There is no deformation left and the resulting classical solution is
precisely the undeformed Hofman-Maldacena solution with $\phi$ replaced by $\alpha.$
The rotation is in the $(\mu_1,\mu_3)$ plane with $\mu_2$ being constant, cf. \eqref{mugen}.
This solution describes the magnon of Eq.~\eqref{HMmagn}. 
However, the $Z$-fields are not given by any single superfield $\Phi_i.$
The rotating field is actually $\varphi^1+i\varphi^5.$
Hence the $\theta-\alpha$ solution corresponds to the magnon \eqref{HMmagn}
with $Z=\varphi^1+i\varphi^5$ and the impurity $I=\Phi_2.$
Of course, in the undeformed $\cN=4$ gauge theory, this magnon is equivalent
to the magnon of the Hofman-Maldacena solution due to the $SO(6)_R$ symmetry.
The same is true in the $\beta$-deformed theory at least at small values of $\beta \ll 1$
relevant for the LM supergravity regime. In this regime the dispersion relation
of the magnon is given by
\eqref{disp}.

\underline{$\gh \neq 0$: String solution parameterized by three angles}

Now we want to study a non-trivial deformation of the Hofman-Maldacena solution
in the $\theta-\phi$ sector.
It will turn out that this solution is required to live on the $\St^3$ sphere.
Hence we need to consider a classical string moving on the $\bigR \times \St^3$.
To achieve this we set $\alpha=0$ and use \eqref{mus3}.
The deformed 3-sphere $\St^3$ is parameterized by the three angles $\theta$, $\phi_1$ and $\phi_2.$
The non-vanishing components of the metric and the two-form field $B^{\rm NS}$ are given by
\bea
ds^2 &=& d\theta^2+G\sin^2\theta\, d\phi_1^2+G\cos^2\theta \, d\phi_2^2\ , \\
B^{\rm NS}_{\phi_1 \phi_2} &=& \hat\gamma \ G \  \sin^2\theta \ \cos^2\theta
\eea

Classical equations follow from the Polyakov action
\bea \label{action}
S=\,-\frac{\sqrt{\lambda}}{2}\int \frac{d\tau dx}{2 \pi}  
\sqrt{-\gamma}\,[\gamma^{\alpha \beta}\partial_{\alpha}X^{\mu} 
\partial_{\beta}X^{\nu}G_{\mu \nu}-\epsilon^{\alpha \beta}\partial_{\alpha}X^{\mu} \partial_{\beta}X^{\nu}B_{\mu \nu}]
\eea
After fixing the gauge through $\gamma^{\alpha \beta}=\eta^{\alpha \beta}=(-1,1)$ and plugging in \eqref{action} 
the expressions for $G_{\mu \nu}$ and $B^{\rm NS}_{\mu \nu}$ we have
\bea
S=\,-\frac{\sqrt{\lambda}}{2}\int \frac{d\tau dx}{2 \pi} \ 
[-(\partial_{\tau}t)^2-(\partial_{\tau}\theta)^2+(\partial_{x}\theta)^2+
G\cos^2\theta((\partial_{x}\phi_2)^2-(\partial_{\tau}\phi_2)^2)\cr +
G\sin^2\theta((\partial_{x}\phi_1)^2-(\partial_{\tau}\phi_1)^2)-
2\hat\gamma G\sin^2\theta\cos^2\theta(\partial_{\tau}\phi_2\partial_{x}\phi_1-
\partial_{x}\phi_2\partial_{\tau}\phi_1)].
\eea
Equations of motion for $t$, $\theta$, $\phi_1$ and $\phi_2$ follow from this 
action.\footnote{We have also solved classical equations which follow from the Nambu-Goto 
action. In this way we found the same solutions and the same expression for the energy as the 
ones written down below. 
This agreement also guarantees that our solutions satisfy  the Virasoro constraints.}
We choose the conformal gauge $t = \tau$ and 
look for the classical solution of the form
\be
\theta(x,t) \, =\, \theta(y) \ , \qquad
\phi_1(x,t) \, =\, t \, +\,  g_1(y) \ , \qquad 
\phi_2(x,t) \, =\,   g_2(y)
\label{s3ans}
\ee
Here $y= cx - dt$ is the same as in the undeformed case. 
The constants $c$ and $d$ are real-valued and positive.
The main difference of the ansatz \eqref{s3ans}
with that in the undeformed case \eqref{HMans} is the appearance of the third angle $\phi_2 = g_2(y).$
It follows from the equations of motion (and in particular from the contributions of the $B^{\rm NS}$ form)
that $\phi_2$ can never be decoupled on the
deformed sphere with $\gh^2 >0.$
This implies that in the $(\theta,\phi_i)$ sector any $\beta$-deformation
of the $S^2$-solution of Hofman and Maldacena will necessarily live on the 3-sphere $\St^3.$ 
Generalizations to motion on higher spheres is straightforward.

\def\hf{{\textstyle{1\over2}}}
The classical equations for our ansatz take the form
\bea
(d^2-c^2)\, \partial^2_y \,\theta  &=& 
\hf\partial_\theta (G\, \sin^2 \theta)\, 
\left(1-2d\, \partial_y g_1 +(d^2-c^2)(\partial_y g_1)^2\right)
\, +\,
\hf \partial_\theta(G\, \cos^2 \theta)\,(d^2-c^2)(\partial_y g_2)^2
\nonumber \\
&&+\,
\partial_\theta(\gh G\, \sin^2 \theta\, \cos^2 \theta)\,c\, \partial_y g_2
\eea
\bea
&&(d^2-c^2)\, \partial_y(G\, \sin^2 \theta\, \partial_y g_1)
\,-\,
d\, \partial_y(G\, \sin^2 \theta) \, =\, 0
\\
&&(d^2-c^2)\, \partial_y(G\, \cos^2 \theta\, \partial_y g_2)
\,-\,
c\, \partial_y(\gh G\, \sin^2 \theta\, \cos^2 \theta)\, =\, 0
\eea
These equations can be simplified as follows
\bea
\partial_y \,g_1 &=& -d\left(1-\frac{1}{G\,\sin^2 \theta}\right) 
\label{g2eq2}\\
\partial_y \,g_2 &=& - \hat\gamma\, c\, \sin^2 \theta
\label{g3eq2}\\
(\partial_y \,\theta)^2 &=& c^2\, \cos^2 \theta\,+\,
d^2\left(1-\frac{1}{G\,\sin^2 \theta}\right)
\label{thetaeq2}
\eea
Here we have imposed $c^2-d^2=1$ which guarantees that the group velocity $v \equiv d/c \le 1.$
We also have applied boundary conditions that 
as $y \to \pm \infty$ the angle $\theta \to \pi/2.$ At the same time the derivatives 
$\partial_y \,\theta$ and $\partial_y\, g_1$ vanish in this limit.
It is easy to see that the derivative of the third 
angle, $\phi_2,$ cannot be vanishing at infinity, $\partial_y \,g_2 \to \,-\gh c \neq 0.$

Substituting the expression \eqref{Gdef} for $G$ 
 into the equation for $\theta$ we get
\be
\partial_y \,\theta \,=\, \cos \theta\, \sqrt{
c^2\,-\, d^2\, \frac{1+\hat\gamma^2 \sin^2 \theta}{\sin^2 \theta}}
\label{thetaeq3}
\ee
This equation can be integrated and admits an analytic solution:
\be
\cos \theta \,=\, \sqrt{\frac{1-\hat\gamma^2d^2}{c^2-\hat\gamma^2d^2}} \,\, 
\frac{1}{{\rm cosh}\,(\sqrt{1-\hat\gamma^2d^2}\,\, y)} \, \equiv\,
\cos \theta_0 \,\, \frac{1}{{\rm cosh}\,(\sqrt{1-\hat\gamma^2d^2}\,\, y)}
\label{thetaansw}
\ee
This expression is reminiscent of the undeformed solution in
\eqref{HMansw}. 
Solutions of the two remaining equations \eqref{g2eq2}, \eqref{g3eq2} can be found 
straightforwardly from the expression for $\cos \theta$ in \eqref{thetaansw}. 
 
We now proceed to evaluate the conserved charges corresponding to the $t$, $\phi_1$ and $\phi_2$ 
isometries of the background. These are the energy $E$ and the two angular momenta 
$J_1$ and $J_2$. They are given by
\bea\label{energy}
E &=& \int_{-\infty}^{\infty}\,  dx\,{\delta S}/{\delta \dot{t}} \, =\, 
\frac{\sqrt{\lambda}}{2\pi}\int_{-\infty}^{\infty}\,  dx\ ,
\\
\label{j2}
J_1 &=& \int_{-\infty}^{\infty}\,  dx\,{\delta S}/{\delta \dot\phi_1} \, =\,
\frac{\sqrt{\lambda}}{2\pi}\int_{-\infty}^{\infty}\, 
dx \  G\sin^2\theta(\partial_{t}\phi_1-\hat\gamma \cos^2\theta\partial_{x}\phi_2)\ ,
\\
\label{j3}
J_2 &=& \int_{-\infty}^{\infty}\,  dx\,{\delta S}/{\delta \dot\phi_2} \, =\,
\frac{\sqrt{\lambda}}{2\pi}\int_{-\infty}^{\infty}\,  
dx \ G\cos^2\theta(\partial_{t}\phi_2+\hat\gamma \sin^2\theta\partial_{x}\phi_1)
\eea
With these expressions we
can now derive the expression for 
the quantity $E-J_1,$ on our solution.
We substitute into 
\eqref{j2} the solutions \eqref{s3ans}-\eqref{g3eq2}, \eqref{thetaansw} and after a 
little algebra we find a simple result
\be
E-J_1\, = \,\frac{\sqrt{\lambda}}{2\pi}\,c^2 \int_{-\infty}^{\infty} \  dx \, \cos^2\theta \, =\,
\frac{\sqrt{\lambda}}{2\pi}\, c\, \int_{-\infty}^{\infty} \,  dy \, \cos^2\theta
\ee

This expression can be rewritten entirely in terms of the angle $\theta$,
\be\label{semifinal}
E-J_1 \,=\, \frac{\sqrt{\lambda}}{\pi}\, c\,\int_{\theta_0}^{\pi/2} \  d\theta \, 
\cos^2\theta\,
(\partial_y \theta)^{-1}
\ee
Using \eqref{thetaeq3}
the integral in \eqref{semifinal} can be performed immediately with the result being
\be
E-J_1 \, =\, \frac{\sqrt{\lambda}}{\pi}\,\sqrt{1+\hat\gamma^2\sin^2\theta_0}\
 \cos\theta_0 \ ,\qquad
0\le \theta_0\le \pi/2
\label{epsansw}
\ee
The expression for the second angular momentum $J_{2}$ can be also simplified in 
a similar fashion giving us the simple result
\be
J_2\,=\, \frac{\sqrt{\lambda}}{\pi}\, \gh\,d\,
\int_{-\infty}^{\infty}\,dy\,  
 \cos^2 \theta\,=\, \hat\gamma\ \frac{d}{c}\ (E-J_1)\ . \label{j2rest}
\ee
For ease of comparing different formulae, it will be useful to express all the answers in terms of $\theta_0.$
It follows from \eqref{thetaansw} that $\sin^2 \theta_0 = d^2/(c^2-\gh^2 d^2).$ This gives the relation for 
$d/c=\sin\theta_0/\sqrt{1+\hat\gamma^2 \sin^2\theta_0}.$ 
This implies that $J_2$ on our classical string solution takes the form
\bea
J_2=\frac{\sqrt{\lambda}}{\pi}\ \hat\gamma\ \sin\theta_0 \ \cos\theta_0.
\label{j3answ}
\eea
This is a remarkable and somewhat surprising result: our embedding of the Hofman-Maldacena giant magnon
solution to the $\beta$-deformed theory has resulted in the appearance of
the second angular momentum $J_2.$
We recall that $J_2$ was identically zero on the original
Hofman-Maldacena solution in the undeformed theory.\footnote{This of course is not inconsistent
with the fact that the Hofman-Maldacena solution corresponds to an elementary magnon of magnon-charge $J_2=1.$  
The vanishing $J_2$ of the classical string only implies that $J_2$ is zero at order-$\sqrt{\lambda},$
where $\lambda \gg 1$ to justify the semiclassical analysis.}
In the deformed case $\gh ={\rm fixed} \neq 0$ 
and thus our solution necessarily
acquires the second angular momentum $J_2.$ The value of $J_2$ in \eqref{j3answ} is proportional to the parameter
$\sqrt{\l} \gh$ which can take any value, large or small. Furthermore, $J_2$ in \eqref{j3answ}
depends on the value of $\theta_0$ which labels different solutions within our ansatz. 
We will clarify the nature of $J_2$ and the interpretation of
the classical string solution as magnon excitations on the spin chain in the next section. 

Since our solution has two angular momenta, $J_1$ and $J_2$ we can ask if its dispersion relation is reminiscent
of \eqref{disp2}.
Remarkably, the dispersion relation is precisely of the square-root form required in
\eqref{disp2}. Using Eqs.~\eqref{epsansw}
and \eqref{j3answ}, we find that on our solution
\be
(E-J_1)^2-J_2^2 \ = \, \frac{\lambda}{\pi^2}\, \cos^2 \theta_0
\ee
Since we want to interpret our solution as a magnon, we will define the
magnon momentum $p$ in terms of the parameter $\theta_0$ 
similarly to the undeformed case \eqref{HMident}
via
\be
\sin \left(\frac{p}{2}-\pi \b\right) \, =\, \cos \theta_0
\label{HMident2}
\ee
Then the dispersion relation is
\be
 E-J_1 \,  =\, \sqrt{J_2^2 \, +\, \frac{\lambda}{\pi^2} \sin^2 \left(\frac{p}{2}-\pi \b\right)}
\label{finfin}
\ee
where $p$ is the momentum carried by the magnon. 
We note that the dispersion relation above depends periodically on $p$ as $p \to p+ 2 \pi,$
and on $\beta$ as $\beta \to \beta + 1$ as required.
In the regime $\lambda \gg 1,$ $E \sim J_1 \to \infty$ and  $J_2$ arbitrary\footnote{Dictated by
\eqref{j3answ}.}
this classical string result is completely reliable.
Essentially, one expects that the only effect of quantum corrections in this regime is the fact that the angular momenta 
are quantized,  see also \cite{MTT}.


%
\section{Interpretation in terms of magnons}

The string solution constructed in section {\bf 2} is a generalisation of the Hofman-Maldacena
solution to the $\beta$-deformed background. We can think of it as the $\beta$- or $\gh$-deformation
of the Hofman-Maldacena classical string. In the limit where the deformation parameter goes to zero, $\gh \to 0,$
our solution collapses to the original Hofman-Maldacena solution, as expected, and can be seen from 
eqs.~\eqref{epsansw},\eqref{j3answ}. The Hofman-Maldacena solution of the undeformed theory carried a single
spin, $J_1$, and was identified in \cite{HM} with an elementary magnon excitation of the spin chain. We have already noted
that the deformed
solution, in addition to $J_1$, carries also a non-zero value of the second spin, $J_2$ given by \eqref{j3answ}. As such,
this deformed solution should describe a magnon excitation of magnon-charge $J_2 \propto \sqrt{\lambda} \gh$ in the $\beta$-deformed theory. 
What happens is that when we start with the elementary magnon described by the Hofman-Maldacena solution in the undeformed theory 
and then turn on the deformation $\gh$ of the background, this induces the charge $J_2$ and the resulting string configuration
corresponds to $J_2$-boundstate of elementary magnons. If this is the case, then we need to 
explain how to construct the elementary magnon in terms of a classical string in the deformed theory. 
This will become clear momentarily.

Note that
the solution in the deformed theory we have studied so far, corresponds to a magnon of charge $J_2$
with a {\it fixed} momentum $p$, such that
$J_2 = (\sqrt{\lambda} \gh /2 \pi)  \sin(p-2\pi \beta).$ 
This is simply a reflection of the fact that our solution describes a minimal deformation of the Hofman-Maldacena solution,
both solutions depend on a {\it single} free parameter,
$\theta_0$ (or $c,d$ with $c^2-d^2=1$, or $p$).
In order to describe magnons with two independent parameters, $J_2$ and $p$, one needs to extend the ansatz
\eqref{s3ans} 
to include dependence on one additional parameter. 
This is easily achieved by looking for a string solution with two spins in the form \cite{AFZ,MTT}
\be
\theta(x,t) \, =\, \theta(y) \ , \qquad
\phi_1(x,t) \, =\, t \, +\,  g_1(y) \ , \qquad 
\phi_2(x,t) \, =\, \nu t \, +\,  g_2(y)
\label{nuans}
\ee
Here $y= cx - dt$ is the same as before with $c^2-d^2=1.$ The new parameter is $\nu$ appearing in the equation for
$\phi_2.$ If one sets $\nu=0,$ eqs.~\eqref{nuans} we are back to the original ansatz \eqref{s3ans} of section {\bf 2}.

This ansatz was used earlier in \cite{AFZ} to obtain string solutions with two spins $J_1,J_2$ in the undeformed theory.
In the paper \cite{BR} which has appeared after the first version of the present paper, the ansatz \eqref{nuans}
was used to study corresponding solutions in the $\beta$-deformed theory. The integrals of motion of the 
$\nu$-extended string solution in the deformed theory were calculated in \cite{BR}:
\be
J_2 \,=\, \frac{\nu c+\gh d}{c}\, (E-J_1)\, , \qquad
E-J_1 \,=\, \frac{\sqrt{\lambda}}{\pi} \frac{c\, \cos^2 \theta_0}{\sqrt{1-(\nu c+\gh d)^2}} \label{ej1nu} 
\ee
where the angular parameter $\theta_0$ is defined in terms of $c,d$ and $\nu$ parameters of the solution via 
\be
\cos^2 \theta_0 \, =\, 
\frac{1-(\nu c+\gh d)^2}{c^2-(\nu c+\gh d)^2}
\label{the0nu}
\ee
These equations generalise expressions in \eqref{j2rest},\eqref{epsansw},\eqref{thetaansw} respectively.
In particular, it follows that expressions for $J_1$ and $\cos^2 \theta_0$ can be obtained from the results of section {\bf 2}
by a shift $\gh d \longrightarrow \nu c + \gh d.$ In what follows it will be convenient to denote this
universal combination as
\be
\Gamma\, := \, \nu c + \gh d
\label{Gammadef}
\ee
We will now demonstrate that the two free parameters of the ansatz \eqref{nuans} can be chosen 
in such a way that the resulting giant magnon solution carries a fixed value of the magnon charge $J_2$
for any value of the magnon momentum $p$. In other words, one can characterise the giant magnon by two
independent arbitrary constant values of $p$ and $J_2$. 
In particular, the value of $J_2$ can be chosen to be one (or more precisely zero in the leading order in $\sqrt{\lambda}$)
to describe the elementary magnon, or different from one to describe a magnon boundstate.

The magnon momentum $p$ is determined via \eqref{HMident2} in terms of $\cos \theta_0$. 
We now fix the value of $p$ (or equivalently of $\cos \theta_0$) and of the spin $J_2$ and solve for the
free parameters of the ansatz in terms of these values. From eqs.~\eqref{ej1nu} we determine the constant 
$\Gamma$ in terms of $J_2$ and $p$ (i.e. $\theta_0$) as
\be
\Gamma^2 \, =\, \frac{J_2^2}{J_2^2 + \frac{\lambda}{\pi^2} \cos^4 \theta_0} \ .
\label{Gamansw}
\ee
Then all the parameters of the ansatz: $c,d$ and $\nu,$ are determined through $\Gamma$ via
eqs.~\eqref{the0nu},\eqref{Gammadef} as
\be
c^2\,=\, \Gamma^2 + \frac{1-\Gamma^2}{\cos^2\theta_0} \ , \quad
d^2 \, =\ c^2-1 \ , \quad 
\nu \, =\, \frac{\Gamma-\gh d}{c} \ .
\label{constansws}
\ee
Thus we have uniquely fixed the two independent parameters of the classical string
in terms of the magnon charge $J_2$ and momentum $p$. The dispersion relation 
still takes the required form \eqref{finfin}.
For $J_2 =1$ the solution describes the elementary magnon, and for $J_2 > 1,$ a magnon
boundstate\footnote{In full quantum theory all angular momenta are quantized. 
Hence when quantum corrections are included, the classical result
for $J_2$ will have to take integer values.} 
in the $\beta$-deformed theory. 
When set the deformation parameter to zero, $\gh \to 0,$
the first equation in \eqref{ej1nu} gives precisely the value of $J_2 \propto \nu $ in the undeformed theory
as expected \cite{AFZ,BR}.

To summarise,
in this paper we have shown that the Hofman-Maldacena construction \cite{HM} of magnons in terms
of classical string solutions can be successfully generalised to $\beta$-deformed theories. 
This generalisation always results in a magnon solution which carries a second orbital momentum $J_2$.
The solution satisfies the exact square-root-type dispersion relation \eqref{finfin}.

\section*{Acknowledgements}   

We acknowledge a useful discussion with Matthias Staudacher.
CSC is supported by EPSRC through an Advanced
Fellowship. VVK and GG are supported by PPARC through a Senior and a Postdoctoral Fellowships.

\end{document}